\documentclass{appolb}
\usepackage{graphicx}
\usepackage{amsmath}
\usepackage{graphicx, graphics, color}
\newcommand{\red}[1]{\textcolor[rgb]{1.00,0.00,0.00}{#1}}

\newcommand{\blue}[1]{\textcolor[rgb]{0.00,0.00,1.00}{#1}}

\newcommand{\be}{\begin{equation}}
\newcommand{\ee}{\end{equation}}
\newcommand{\beq}{\begin{eqnarray}}
\newcommand{\eeq}{\end{eqnarray}}

\graphicspath{{./figs/}}

\begin{document}
\title{Beyond Hubbard: the role of correlated hopping interaction in superconductors and quantum dot devices%
\thanks{Presented at "The Concepts in Strongly Correlated Quantum Matter Conference (CSCQM)", Krak\'ow,  November 20 to 22, 2025}%
}
\author{Karol I.\ Wysoki\'{n}ski
\address{Institute of Physics, M.~Curie-Sk{\l}odowska University, pl.~M.~Curie-Sk{\l}odowskiej 1, 20-031 Lublin, Poland\\
https://orcid.org/0000-0002-5366-4455 
}
\\[3mm]
Marcin M. Wysoki\'nski
\address{International Research Centre MagTop, Institute of
   Physics, Polish Academy of Sciences,\\ Aleja Lotnik\'ow 32/46,
   PL-02668 Warsaw, Poland} 
	https://orcid.org/0000-0002-7881-5168
   } 
\maketitle
\begin{abstract}
We investigate the role of strong Coulomb interactions beyond the standard Hubbard model in two distinct physical contexts. First, we analyze the superconducting phase transition occurring near the Mott metal–insulator transition. Second, we study transport properties of artificial nano-scale structures containing quantum dots coupled to external electrodes. In both cases, we focus on the impact of the correlated (assisted) hopping (CH) interaction.
For superconductors, CH acts as a driving mechanism for the phase transition and modifies the spectral properties of the system. We present the evolution of the spectral function as the system approaches the Mott-type transition under varying model parameters.
In quantum-dot-based devices, CH influences the tunneling amplitude between the dot and metallic leads. We demonstrate that the characteristic changes in the conductance of a normal metal–quantum dot–normal metal structure provide a clear signature of the presence and sign of CH interaction.
\end{abstract}

\section{Introduction}

Understanding the role of strong interactions in matter is an old and important scientific problem~\cite{hubbard1963} that has gained wide recognition in the condensed matter community~\cite{spalek2023,spalek2022}, especially after the discovery of high-temperature superconducting oxides~\cite{bednorz1986}. Strong repulsion between electrons has been considered an important factor in the proper description of this family of superconductors~\cite{anderson1987}, or even the main source of pairing. In narrow-band metals in which the interaction strength is of the order of the bandwidth, the local Hubbard term of the type $Uc^{\dagger}_{i\uparrow}c_{i\uparrow}c^{\dagger}_{i\downarrow}c_{i\downarrow}$, in Wannier representation, is certainly the most important. However, besides this one, there exist other longer-range interactions~\cite{hubbard1963,spalek2023} which also play a role. In particular, the interaction of the type $K_{ij}c^\dagger_{i\sigma}c_{j\sigma}(n_{i\bar{\sigma}}+n_{j\bar{\sigma}})$ involving two neighbouring sites {\it i} and {\it j} is of our interest. It is called correlated hopping (CH) or assisted hopping~\cite{micnas1989} and sometimes the charge-bond interaction~\cite{appel1993}. CH has been shown to lead to the superconducting instability of the normal state~\cite{micnas1989} and has been proposed as a proper model of high-temperature  superconductors ~\cite{hirsch1989}. There  exists  an extensive literature on the effect of the correlated hopping interaction on the properties of materials and in particular superconductors~\cite{domanski1994,penc1994,bulka1998,schmidt2006,hubsch2006,schmidt2008,filho2009,Mizia2011,zegrodnik2017,mmwysokinski2017}. This interaction term breaks electron-hole symmetry and thus might be partially responsible for the asymmetry of the superconducting domes between electron- and hole-doped cuprates~\cite{mmwysokinski2017,armitage2010} and other properties~\cite{kiw1992,domanski1993}.

It has also been recognised~\cite{glazman1988,ng1988} that the strong (Hubbard) interaction is of primary importance in nano-structures containing quantum dots tunnel-coupled to external electrodes. In such circumstances, the repulsion between opposite-spin electrons on a dot induces the Kondo correlations~\cite{goldhaber-gordon1998}, leading at low temperature to increased conductance across the structure.  
Remarkably, the importance of the correlated hopping interaction in nanostructures can be easily seen on physical grounds. Imagine the structure consisting of a quantum dot (QD), which is considered here as a small grain with a quantised electron spectrum. Due to the small size and small capacitance of the dot, there is a strong repulsion between the electron occupying it and the second electron which tunnels onto it. This repulsion is a price one has to pay to have two electrons on a dot. This ``charging energy'' is just the Hubbard local interaction $U$. Now, if an electron from one of the external metallic electrodes to which the QD is coupled hops/tunnels onto it, it is obvious that the amplitude of the process depends on whether the QD is empty or occupied. If the relevant state on the dot is occupied by a spin $\sigma$ electron, the Pauli principle allows only an electron with the opposite spin to tunnel. Simple electrostatics suggests that the hopping amplitude onto the dot differs if it is occupied or empty. This shows that CH is an interaction that is always present in systems with quantum dots. The effect of the CH on the transport properties of a normal metal - quantum dot - normal metal (N-QD-N) structure is our second goal.  

In the next Section, the theory of superconductivity in metals close to the Mott transition is formulated, while in Section \ref{sec:QD}, the effect of the correlated hopping on transport in a normal metal - QD - normal metal (N-QD-N) structure has been discussed. We end up with the summary and conclusions in Section \ref{sec:sum}.

\section{Correlated hopping induced superconductivity in strongly correlated metal}\label{sec:sc}

The most famous class of unconventional superconductors, which in the normal state are on the verge of an interaction-induced metal-insulator transition, are copper oxides, better known as high-temperature superconductors (HTS). It means that their description requires a model featuring a Mott transition.
Based on that observation, Phillips {\it et al.}~\cite{phillips2020} proposed to use the Hatsugai-Kohmoto model~\cite{hatsugai1992} for the description of the normal state of the superconducting system. The model is like a standard Hubbard model but with an onsite interaction diagonal in k-space
\begin{equation}
H=\sum_{k\sigma}\xi_{k\sigma}c^\dagger_{k\sigma}c_{k\sigma}+\sum_k U_k n_{k\uparrow}n_{k\downarrow}.
\label{SHK}
\end{equation}
It is appropriate to recall that the model (\ref{SHK}) has been discussed already in 1988 by Spa\l{}ek and W\'ojcik~\cite{spalek1988} in the context of a spin-liquid state. The statistical properties of the model were elaborated later by Byczuk and Spa\l{}ek~\cite{byczuk1994}. Based on those facts, we think that the more proper name for the model would be the Spa\l{}ek-Hatsugai-Kohmoto (SHK) model, the name and acronym we shall use in the following. Similar models were studied in the literature~\cite{baskaran1991,tarasewicz2009,jablonowski2023} as reviewed recently~\cite{zhao2025}.

In Eq. (\ref{SHK}) we slightly generalised~\cite{yang2021} the original version~\cite{hatsugai1992,byczuk1994} of the model by  allowing for the wave-vector dependence of the interaction parameter $U$ and including the chemical potential $\mu$ into $\xi_{k\sigma}=\epsilon_{k\sigma}-\mu$. The k-dependent interaction in the form $U_k=U-2T_x\cos(k_xa)-2T_y\cos(k_ya)$ has been found~\cite{yang2021} to lead to the celebrated Fermi arcs observed in some unconventional superconductors and topological materials, like $e.g.$ Weyl semimetals. Even more complicated $k$ dependences were proposed and studied~\cite{wysokinski2023}. However, for all calculations in this work we have taken $U_k=U$. We also allowed for a spin dependence of the single-particle energies, having in mind studies of  $e.g.$  the ferromagnetic superconductors. 

It turns out that the SHK model (\ref{SHK}), which can be exactly solved, describes the Mott transition, and moreover, when complemented by the interaction leading to superconducting instability, allows studying the superconducting state across the metal-insulator transition. The space- and frequency-dependent normal state GF, in Zubarev notation~\cite{zubarev1960}, defined as $G_\sigma(k\omega)=\langle\langle c_{k\sigma}|c^\dagger_{k\sigma}\rangle\rangle_\omega$ is easily derived  
\be  
G_\sigma(k\omega)=\frac{\omega-\xi_{k\sigma}-U_k(1-\langle n_{k\bar{\sigma}}\rangle)}{(\omega-\xi_{k\sigma})(\omega-\xi_{k\sigma}-U_k)}=\frac{1-\langle n_{k\bar{\sigma}}\rangle}{\omega-\xi_{k\sigma}}+\frac{\langle n_{k\bar{\sigma}}\rangle}{\omega-\xi_{k\sigma}-U_k}.  
\ee  
The single-particle energy of a 3d cubic crystal equals $\epsilon_{k\sigma}=-2t\gamma(k)$, where $t$ is the nearest-neighbour hopping amplitude and $\gamma(k)=\cos(k_xa)+\cos(k_ya)+\cos(k_za)$ and $a$ denotes the lattice constant. In the following, we set $t=1$ as our energy unit, and mainly work with units in which $\hbar=1$.

Before proceeding with the description of the superconducting state, we comment on a few features inherent to the SHK model. In the Hamiltonian (\ref{SHK}), there is no mixing between the lower and upper Hubbard band~\cite{phillips2020}. This means that the excitations in the lower Hubbard band with energy $\xi_{k\sigma}$ are created by the operators $c_{k\sigma}(1-n_{k\bar{\sigma}})$, while excitations in the upper band with energy $(\xi_{k\bar{\sigma}}+U_k)$ are created by the operator $c_{k\sigma}n_{k\bar{\sigma}}$ acting on the doubly occupied states. This is an important aspect of strongly correlated systems. Within the real-space Hubbard model, analogous excitations acting on singly and doubly occupied states have also been considered~\cite{domanski1994} in the context of the superconducting transition in HTSs induced by correlated hopping in the presence of strong Hubbard repulsion.

To study the superconducting transition in  the  Mott state, one has to supplement the model (\ref{SHK}) by a term leading to the superconducting state. We add the correlated hopping term, which in real space reads
\begin{equation}
H_{CH}=\sum_{ij\sigma}K_{ij}(c^\dagger_{i\sigma}c_{j\sigma}+c^\dagger_{j\sigma}c_{i\sigma})(n_{i\bar{\sigma}}+n_{j\bar{\sigma}}).
\end{equation}
In the above, the indices $i,j$ denote lattice sites in a crystal and play a role of quantum numbers, like the wave vector $k$ does in reciprocal space. The BCS-reduced version of $H_{CH}$, for interaction between nearest-neighbour sites $i$ and $j$ being constant $K_{ij}=K$, in wave-vector space is written as
\begin{equation}
H^{BCS}_{CH}=\sum_{k,k'}V_{k,k'}c^\dagger_{k\uparrow}c^\dagger_{-k\downarrow}c_{-k'\downarrow}c_{k'\uparrow}\approx\sum_k [\Delta_k c^\dagger_{k\uparrow}c^\dagger_{-k\downarrow}+\Delta^*(k)c_{-k\downarrow}c_{k\uparrow}].
\label{BCS-CH}
\end{equation}
and is treated in a mean-field-like fashion. In equation (\ref{BCS-CH}), we denoted~\cite{hirsch1989} 
\be
V_{k,k'}=-\frac{K}{2}(\gamma(k)+\gamma(k')).
\label{vkk-red}
\ee
The gap function $\Delta_k$ is given by
\be
\Delta_k=\sum_{k'}V_{k,k'}\langle c_{-k'\downarrow}c_{k'\uparrow}\rangle,
\label{def-delta}
\ee
and, as visible from (\ref{def-delta}) and (\ref{vkk-red}), has a general k-dependence $\Delta_k=\Delta_0+\Delta_1\gamma(k)$, with both coefficients depending on temperature~\cite{micnas1989,hirsch1989}. To calculate the average $\langle c_{k\uparrow}c_{-k\downarrow}\rangle$ for a Hamiltonian with $k$-space interaction, we shall use the Green function method (in Gorkov-Nambu representation) and the fluctuation-dissipation theorem, which relates the said average to the frequency - dependent retarded (r) Green function
\be
\langle c_{-k\downarrow}(t)c_{k\uparrow}(t)\rangle=\int{\frac{d\omega}{2\pi} \langle c_{-k\downarrow}c_{k\uparrow}\rangle_\omega}=\int d\omega f(\omega)\frac{-1}{\pi}Im\langle\langle c_{k\uparrow}|c_{-k\downarrow}\rangle\rangle^r_\omega, 
\label{dis-fluct}
\ee
where $f(\omega)=(e^{\frac{\omega}{k_BT}}+1)^{-1}$ is the Fermi-Dirac distribution function, $k_B$ is the Boltzmann constant and $T$ is the temperature.

To describe  the  superconducting state, one needs the Green function (GF) in $\omega$- and $k$-space written as a two-by-two matrix 
\begin{align}
G(k,\omega)=\left( \begin{array}{lr}
 \langle\langle c_{k\uparrow}|c^\dagger_{k\uparrow}\rangle\rangle_\omega & \langle\langle c_{k\uparrow}|c_{-k\downarrow}\rangle\rangle_\omega \\
\langle\langle c^\dagger_{-k\downarrow}|c^\dagger_{k\uparrow}\rangle\rangle_\omega &\langle\langle c^\dagger_{-k\downarrow}|c_{-k\downarrow}\rangle\rangle_\omega
    \end{array}\right). 
\label{gkomega}		
\end{align}
Using the equation of motion to find the GF (\ref{gkomega}), one generates the whole series of higher-order Green functions. In general, the technique does not allow one to find a closed solution. However, for the model at hand, the exact solution can be obtained. It involves twelve different Green functions. For the sake of simplicity and clarity, we denote them by the Latin letters from $\underline{A}$ to $\underline{M}$, with $G(k,\omega)\equiv\underline{A}$, and neglect the arguments.  
Thus, the first equation for GF $G(k,\omega)\equiv\underline{A}$ reads 
\begin{align}
\left( \begin{array}{cc}
\omega-\xi_{k\uparrow}&-\Delta_k \\
-\Delta^*_k&\omega+\xi_{-k\downarrow}
\end{array} \right) \underline{A}=
\left( \begin{array}{cc}
1&0 \\
0&1\end{array} \right)+\left(\begin{array}{cc} 
U_k&0 \\
0&-U_{-k} \end{array}\right)\underline{B},
\end{align} 
and involves a new Green function $\underline{B}$.  
At this point, it is useful to have a closer look at the operator structure of the novel GF $\underline B$. This function contains creation (annihilation) operators multiplied by the appropriate particle number operators and describes processes in which Cooper pairs are formed if the state contains additional particles described by additional operators
\begin{align}
\underline{B}=
\left( \begin{array}{cc}
\langle\langle c_{k\uparrow}n_{k\downarrow}|c^\dagger_{k\uparrow}\rangle\rangle_\omega &\langle\langle c_{k\uparrow}n_{k\downarrow}|c_{-k\uparrow}\rangle\rangle_\omega\\
\langle\langle c^\dagger_{-k\downarrow}n_{-k\uparrow}|c^\dagger_{k\uparrow}\rangle\rangle_\omega&\langle\langle c^\dagger_{-k\downarrow}n_{-k\uparrow}|c_{-k\downarrow}\rangle\rangle_\omega
\end{array} \right).
\end{align}
Note that, like in GF $\underline{B}$, also in all higher-order Green functions, the second operators in each entry are the same as in (\ref{gkomega}). The left-hand-side operators in the second row can be obtained from those in the first row by Hermitian conjugation and the changes ($k\leftrightarrow -k$) and ($\uparrow \leftrightarrow \downarrow$). Having this in mind, it is enough to know the $11$ element of any of the Green functions to reproduce the whole matrix.

Before writing the set of equations, it is convenient to define a few auxiliary matrices
\begin{align}
\underline{U}=
\left( \begin{array}{cc}
U_k&0\\
0&-U_{-k}
\end{array} \right);
\underline{\Delta_1}=
\left( \begin{array}{cc}
0&\Delta_k  \\
\Delta^*_k&0
\end{array} \right);
\underline{\Delta_2}=
\left( \begin{array}{cc}
\Delta_{-k}&0\\
0&-\Delta^*_{-k}
\end{array} \right),
\end{align} 
and
\begin{align}
\underline{C_0}=
\left( \begin{array}{cc}
\omega-\xi_{k\uparrow}&0\\
0&\omega+\xi_{-k\downarrow}
\end{array} \right).
\end{align} 

With that notation, we define the matrices $\underline{A_0}=\underline{C_0}-\underline{\Delta_1}$ and $\underline{B_0}=\underline{C_0}-\underline{U}$.

Continuing the process of calculating novel GFs, we shall encounter entries with a more complicated set of operators. Applying the equation of motion technique to these novel functions leads to the following set of equations
\begin{eqnarray}
\underline{A_0}\cdot\underline{A}&=&{\bf 1}+\underline{U}\cdot\underline{B},\\ 
\underline{B_0}\cdot\underline{B}&=&\underline{B_1}+\underline{\Delta_1}\cdot\underline{C}+\underline{\Delta_2}\cdot\underline{D}-\underline{\Delta^*_2}\cdot\underline{E}, \label{eq14}\\
\underline{C_0}\cdot\underline{C}&=&\underline{C_1}+\underline{U}\cdot\underline{F}+\underline{\Delta_2}\cdot\underline{D}-\underline{\Delta^*_2}\cdot\underline{E},\label{eq15}\\
\cdots& & \nonumber
\end{eqnarray}
and so on until the last item $\underline{M}$, which contains only earlier found functions.  The {\bf 1} above denotes the unit matrix, and other novel matrices are defined as
\begin{equation}
\underline{B_1}= 
\left( \begin{array}{cc}
\langle n_{k\downarrow}\rangle&0\\
0&\langle n_{-k\uparrow}\rangle
\end{array} \right);\quad
\underline{C_1}=
\left( \begin{array}{cc}
\langle n_{-k\uparrow}\rangle&0\\
0&\langle n_{k\downarrow}\rangle
\end{array} \right).
\end{equation}

Typically, the infinite chain of equations is cut by projecting some higher-order function onto the lower ones. Here, we deal with a finite set of twelve equations and, in principle, we may solve them without approximations. This, however, is complicated and beyond the scope of this paper. The lowest nontrivial approximation is enough for our purposes, as the system describes the superconducting state across Mott transition. Looking at the above set of equations, we observe that GF $\underline{C}$, which has a structure
\begin{align}
\underline{C}=
\left( \begin{array}{cc}
\langle\langle c_{k\uparrow}n_{-k\uparrow}|c^\dagger_{k\uparrow}\rangle\rangle_\omega &\langle\langle c_{k\uparrow}n_{-k\uparrow}|c_{-k\uparrow}\rangle\rangle_\omega\\
\langle\langle c^\dagger_{-k\downarrow}n_{k\downarrow}|c^\dagger_{k\uparrow}\rangle\rangle_\omega&\langle\langle c^\dagger_{-k\downarrow}n_{k\downarrow}|c_{-k\downarrow}\rangle\rangle_\omega
\end{array} \right)
\end{align} 
can be projected onto $\underline{A}$, with the effect $\underline{C}\approx\underline{C_1}\cdot\underline{A}$. Importantly, similar approximations applied to GFs $\underline{D}$ and $\underline{E}$ make their combination in Eqs.(\ref{eq14}) and (\ref{eq15}) vanish, and the resulting closed system is easily solved for $G(k,\omega)$. In this paper, we are content with that approximation, which describes both, the Mott transition and the superconductivity. The effect of interactions $U$ on the superconducting state is described in a mean-field-like fashion, while the system undergoes Mott transition with increasing $U$.

\begin{figure}[h]
\includegraphics[width=0.32\linewidth]{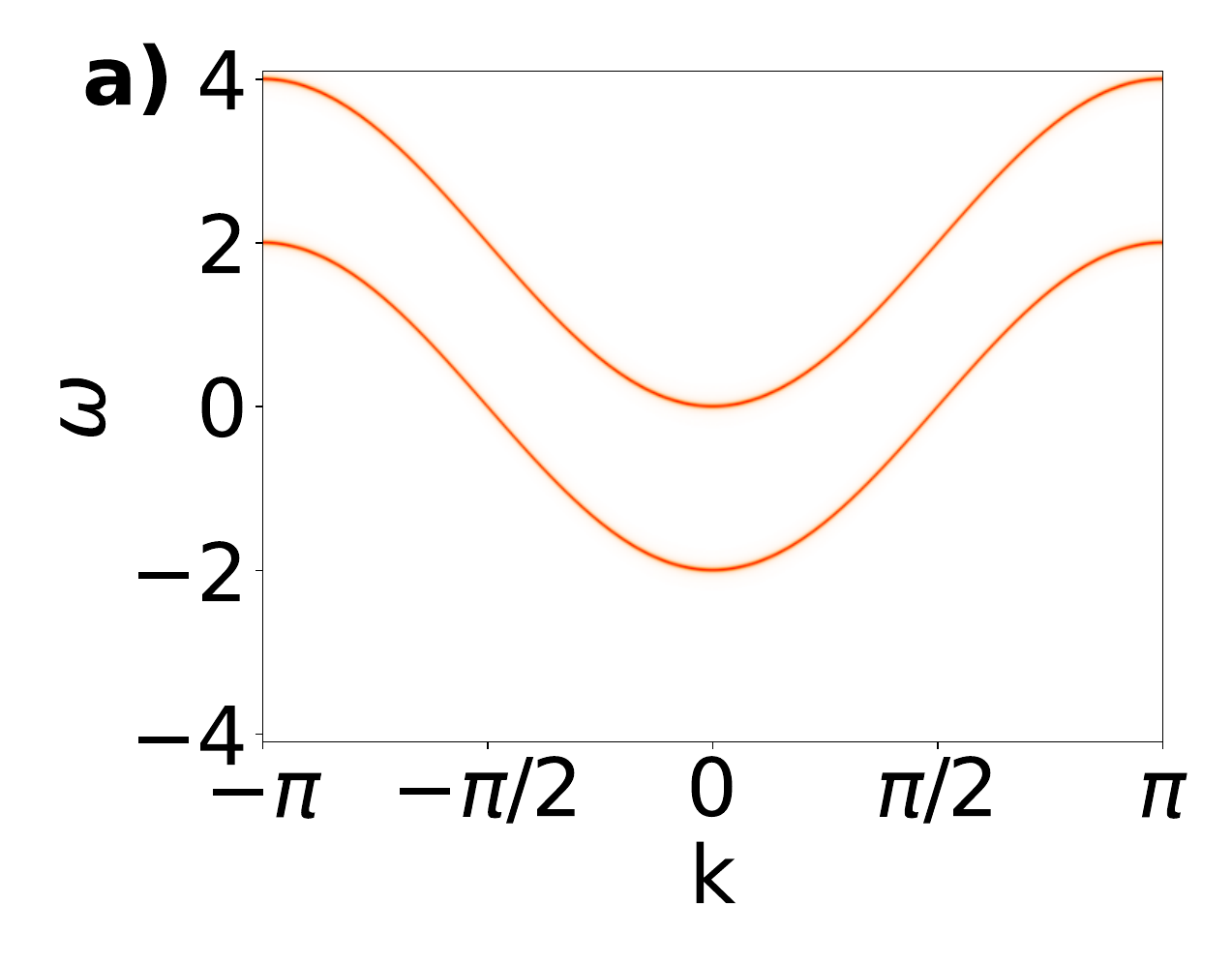}
\includegraphics[width=0.32\linewidth]{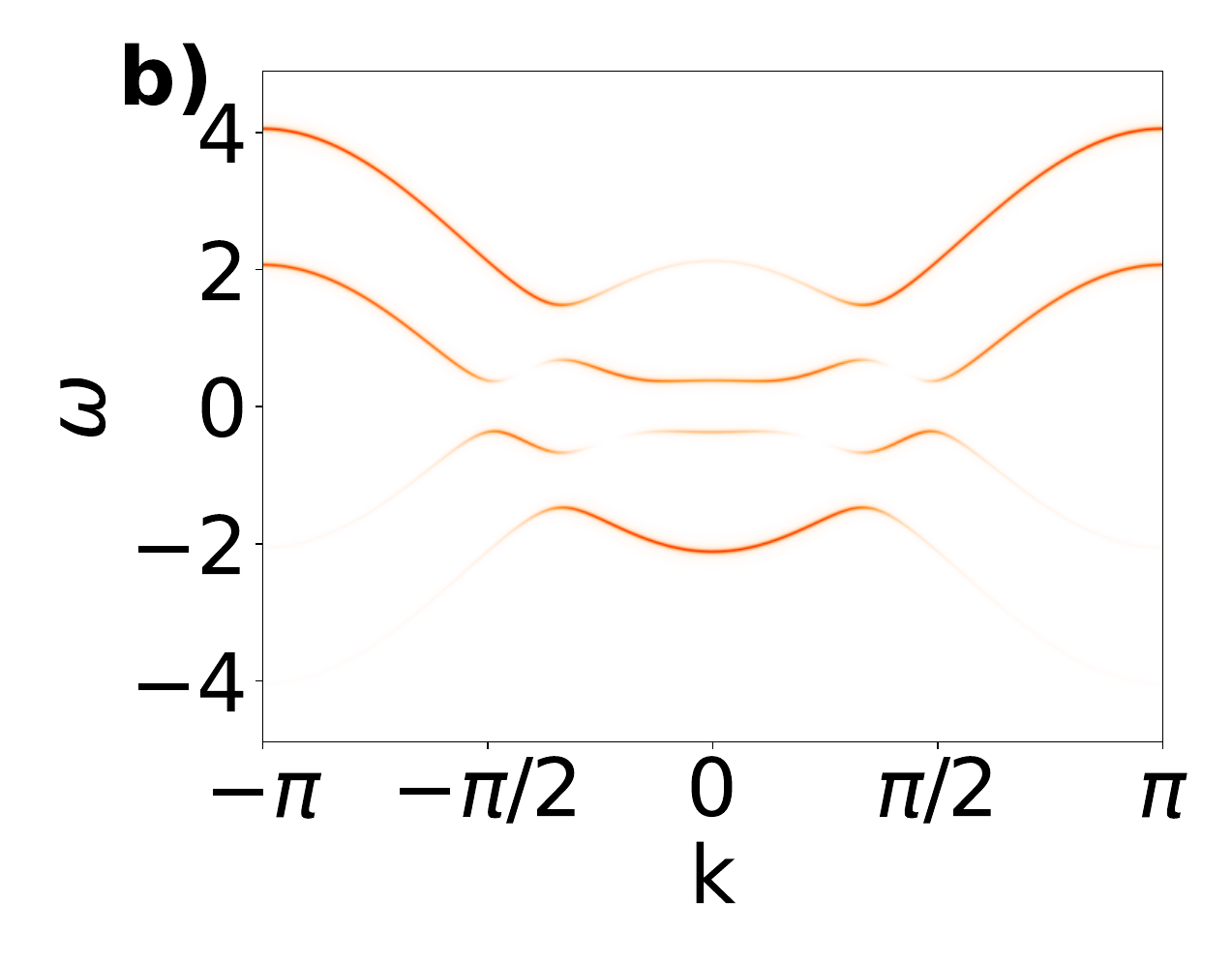}
\includegraphics[width=0.32\linewidth]{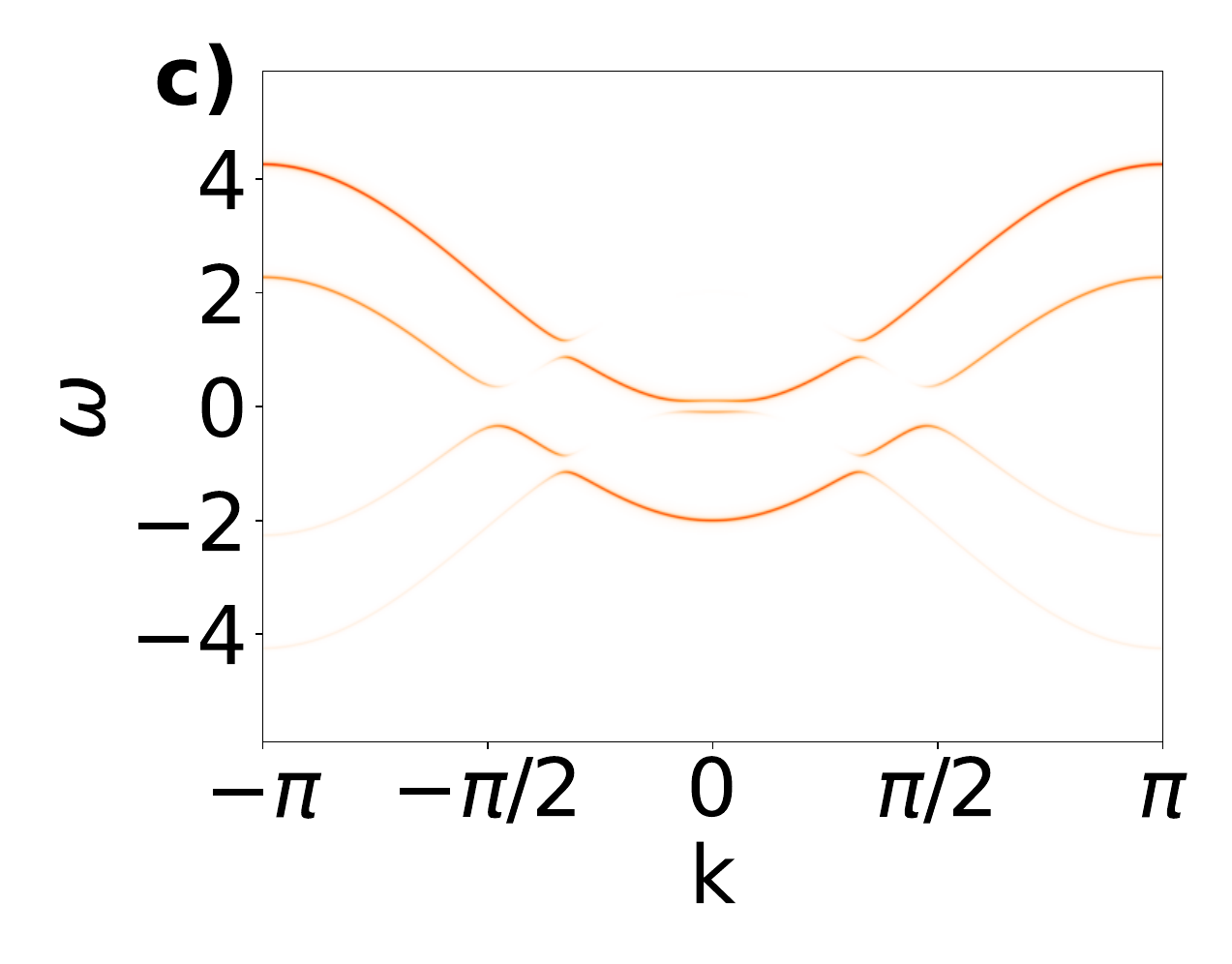} 
\caption{Spectral functions of  (a) normal metal calculated for $U=2$, (b) s-wave superconductor with constant order parameter $\Delta_0=0.8$  and (c) correlated hopping superconductor with $\Delta(k)=\Delta_0+\Delta_1 \gamma(k)$ with $\Delta_1=0.5$. Comparison of the middle and right figures illustrates the changes of the spectral function induced by the particle-hole symmetry breaking term $\Delta_1\gamma(k)$.}
\label{fig:rys1}  
\end{figure} 

\begin{figure}[h]
\includegraphics[width=0.32\linewidth]{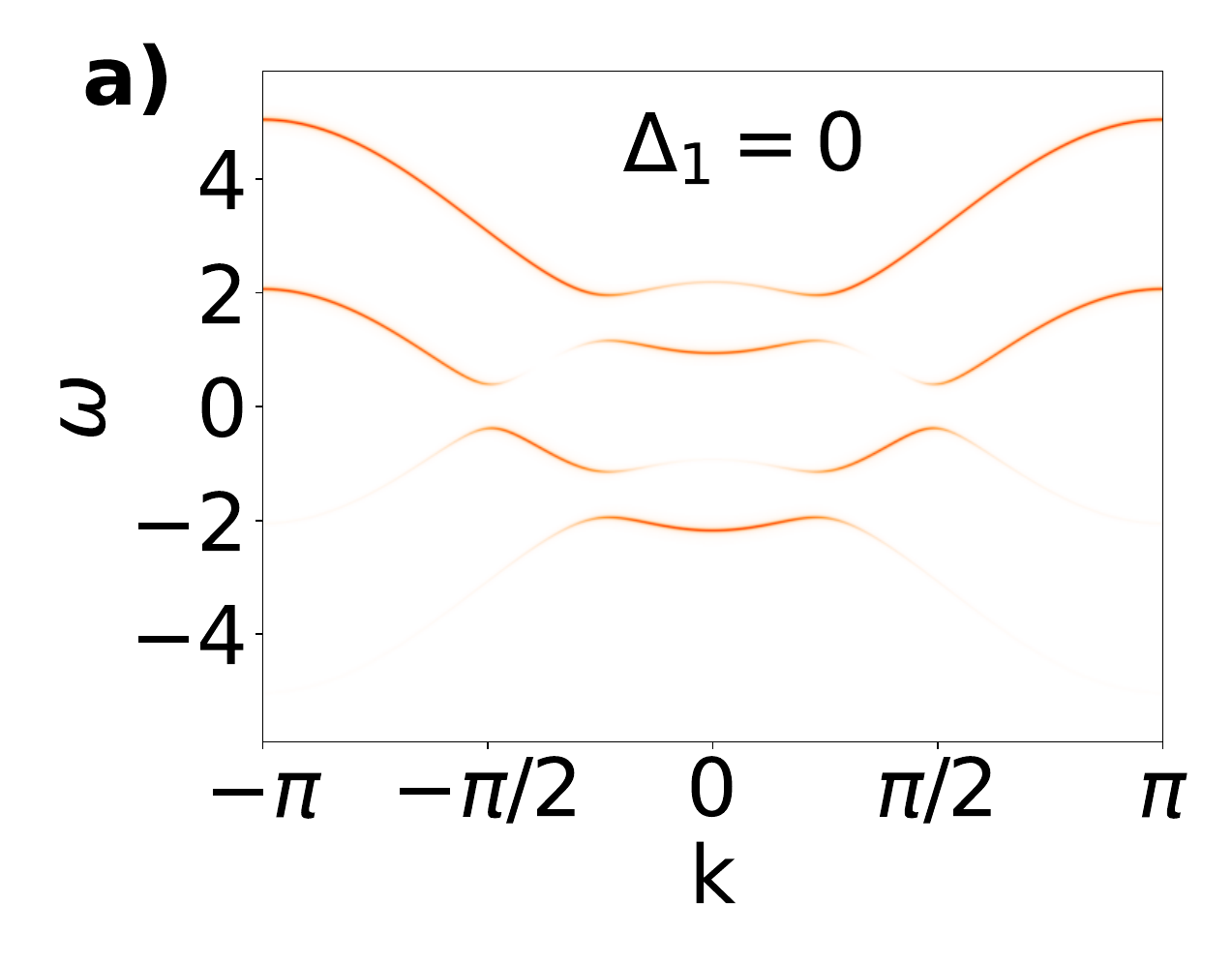}
\includegraphics[width=0.32\linewidth]{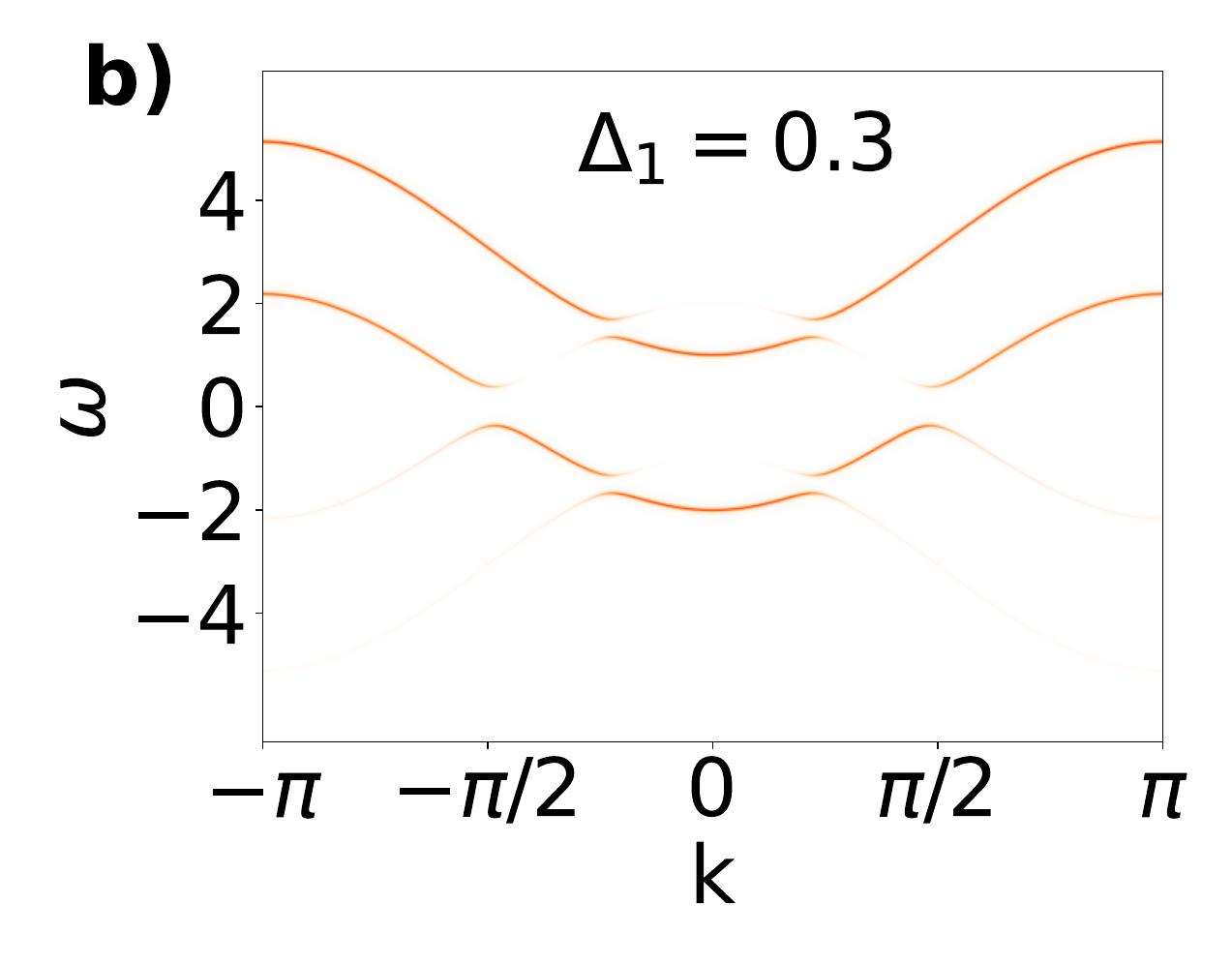}
\includegraphics[width=0.32\linewidth]{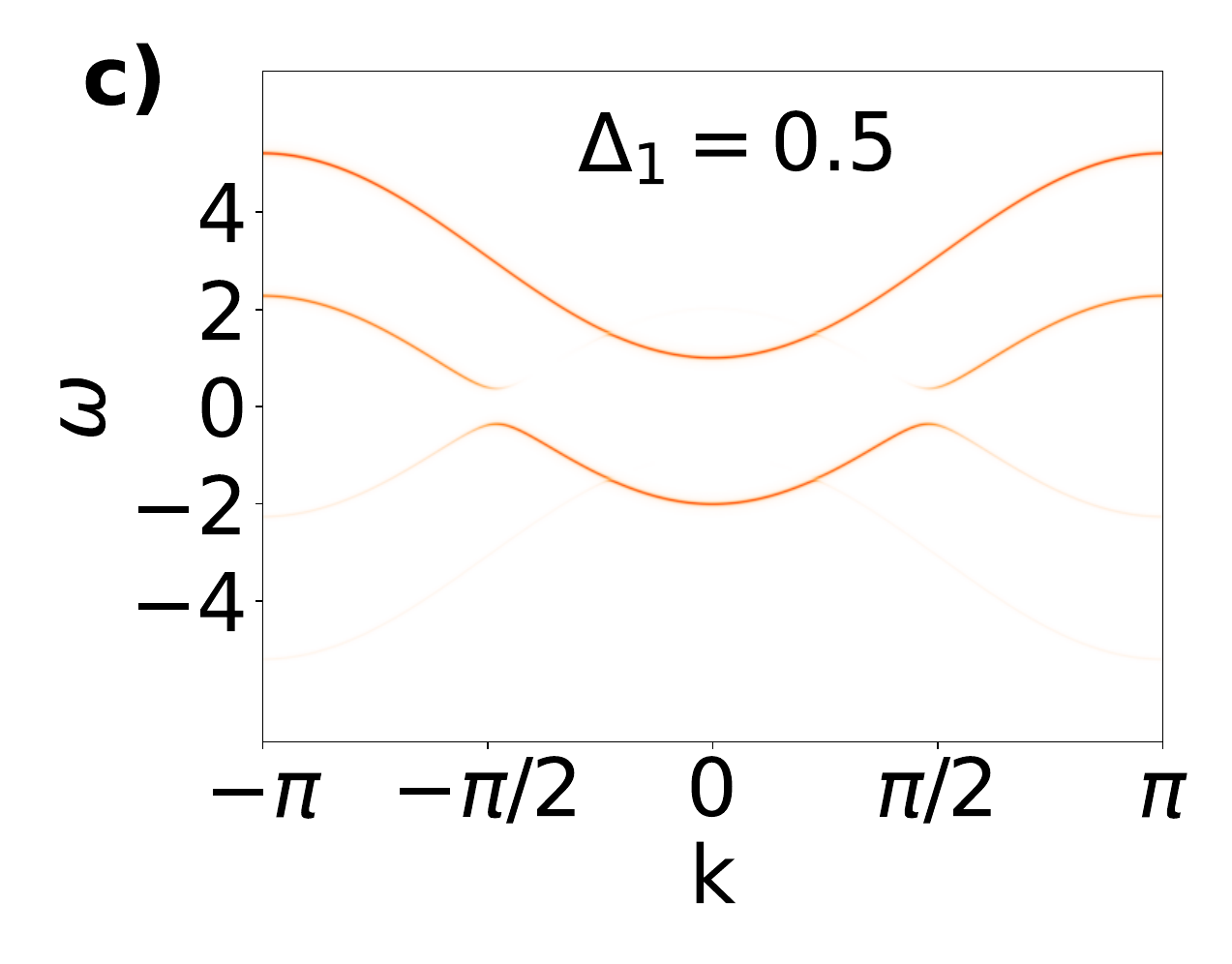}
\caption{Evolution of the spectral function calculated for $U=3$ and $\Delta_0=0.8$, $\Delta_1=0.0$ (a), $\Delta_0=0.8$, $\Delta_1=0.3$ (b) and $\Delta_0=0.8$, $\Delta_1=0.5$ (c). 
}
\label{fig:rys1a}  
\end{figure}

\begin{figure}[h]
\includegraphics[width=0.32\linewidth]{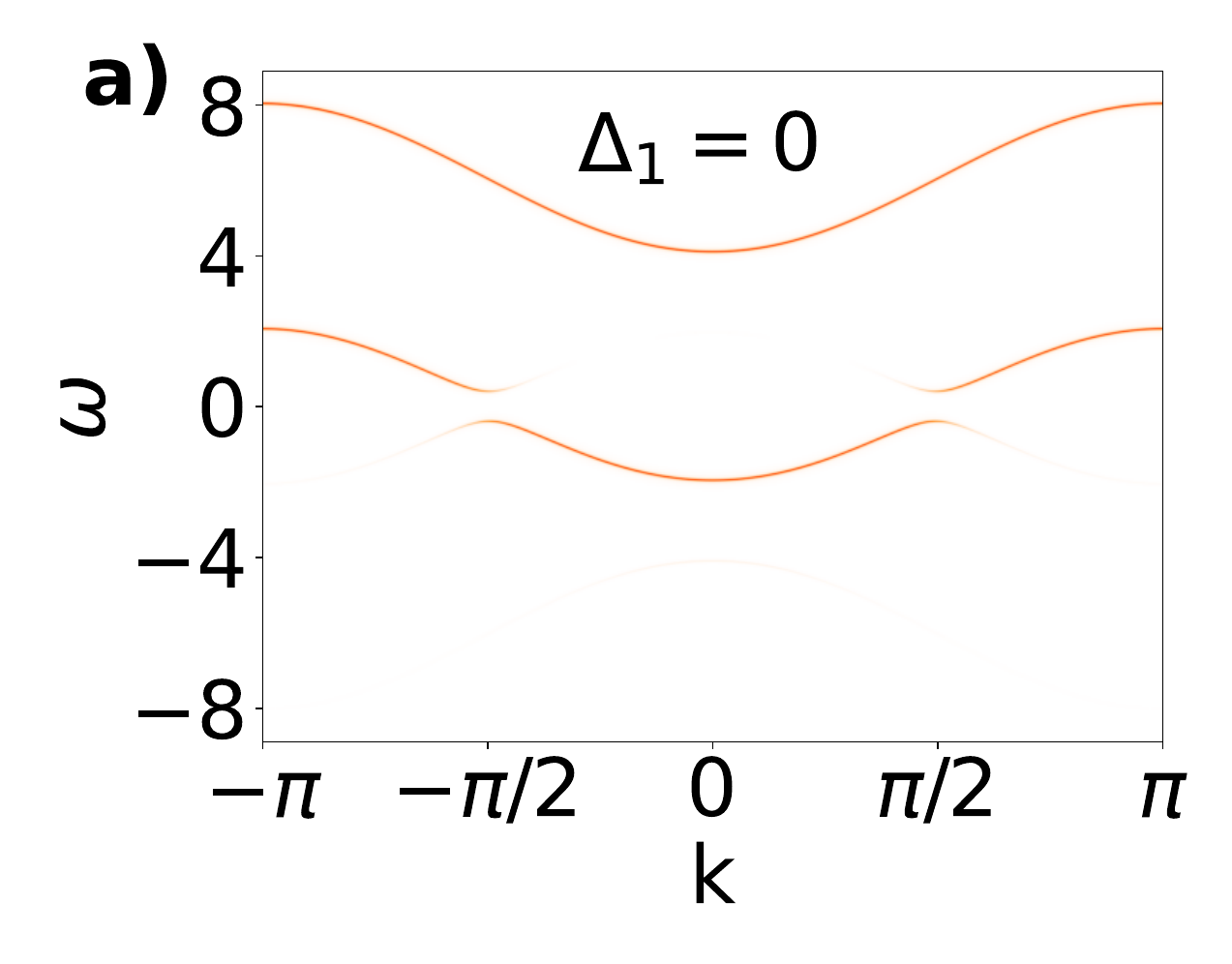}
\includegraphics[width=0.32\linewidth]{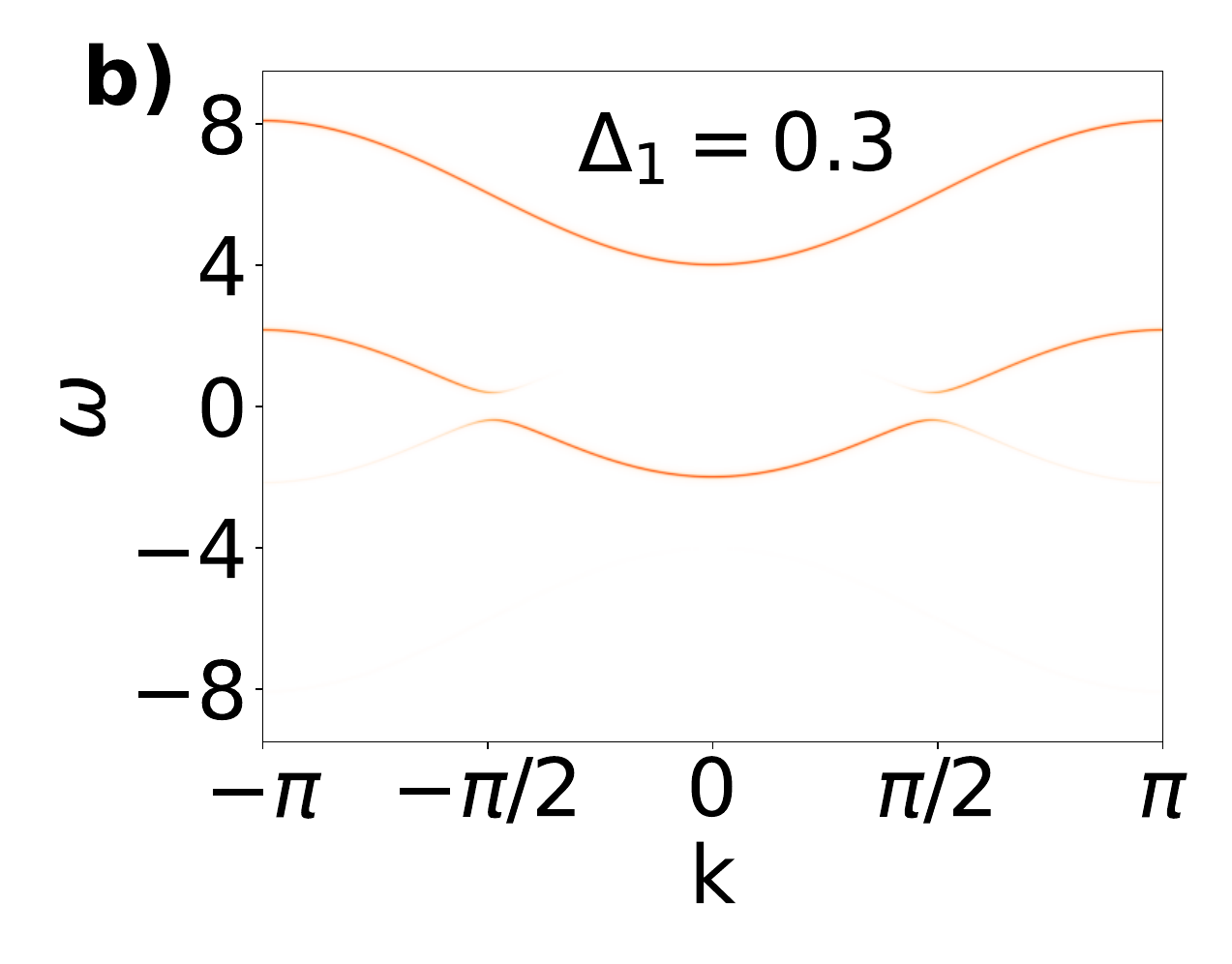}
\includegraphics[width=0.32\linewidth]{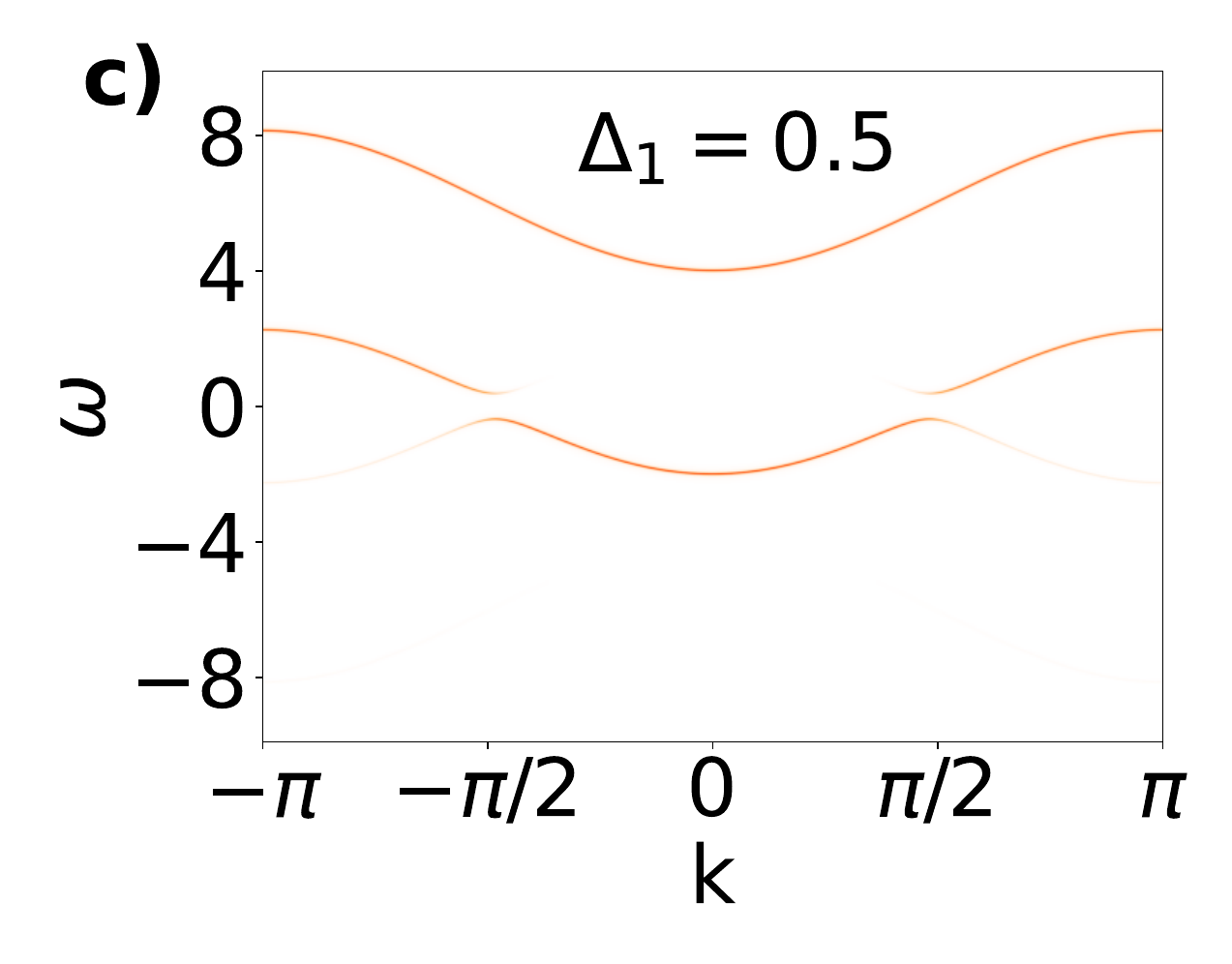}
\caption{Spectral function calculated for s-wave superconductor in doped Mott insulator: $U=6$ and $\Delta_0=0.8$, $\Delta_1=0.0$ (a), $\Delta_0=0.8$, $\Delta_1=0.3$ (b) and $\Delta_0=0.8$, $\Delta_1=0.5$ (c). The magnitude of the superconducting gap at the Fermi level crossing the lower Hubbard band is nearly independent of $\Delta_1$. 
}
\label{fig:rys1b}  
\end{figure} 
To illustrate the results, we present the  spectral function, which is defined as $-Im\, G(k,\omega+ i0^+)/\pi$. In Fig. (\ref{fig:rys1}), spectral function is shown for the correlated system with $U=2$, well below the Mott transition. We have assumed $\langle n_{-k\uparrow}\rangle=\langle n_{k\downarrow}\rangle =0.5$, and put the Fermi level at $\mu=0$. For illustration purposes, we have taken an electron spectrum of a 1-dimensional lattice with $\gamma(k)=-2\cos(k_xa)$. The left or $a)$ panel of Fig.  (\ref{fig:rys1}) refers to the normal state. Two interaction-split bands are separated by $U$.  The upper subband touches the Fermi level (at $\mu=0$), while the lower one is half-filled.

Before presenting the spectral function in the superconducting state, an explanation is needed. With CH interaction (\ref{vkk-red}) and definitions (\ref{def-delta}) and $\Delta_k=\Delta_0+\Delta_1\gamma(k)$, one has to solve the resulting system of equations to find the temperature and carrier concentration dependent parameters $\Delta_0$ and $\Delta_1$ in a self-consistent manner. Here, instead, we shall choose arbitrary values for both of these parameters based on our earlier experience~\cite{mmwysokinski2017}, albeit for a different repulsive interaction. This lack of consistency introduces only small quantitative changes in the resulting spectral functions, and 
is good enough for the purpose of illustrating the effect the CH has on the superconductor's spectral function.

The  panel $b)$ of Fig. (\ref{fig:rys1}) presents the spectral density of the superconductor with pure s-wave order parameter $\Delta_k=\Delta_0=0.8$. One observes the gap function extending from nearly $-\Delta_0/2$ to $\Delta_0/2$. In fact, it extends over a slightly smaller range. In the right panel (Fig. (\ref{fig:rys1}c), the CH-interaction-induced superconductivity with $\Delta_k=\Delta_0+\Delta_1\gamma(k)$ (so called s*-symmetry) is shown for the same value of $\Delta_0=0.8$ and for $\Delta_1=0.5$. One observes significant changes in the magnitude of the gap at different parts of the one-dimensional Brillouin zone (BZ). The gap near the BZ centre is very small, of the order of 0.1, while that close to $\pi/2$ is about $0.3$. Darker colours of the lines indicate higher intensity, which also changes with the gap parameters.

In Fig. (\ref{fig:rys1a}), we show the changes in the spectral function with $\Delta_1$ of the superconductor described by $U=3$ and $\Delta_1=0, 0.3, 0.5$ in the lect (Fig. \ref{fig:rys1a}a), middle (Fig. \ref{fig:rys1a}ab), and right (Fig. \ref{fig:rys1a}c) panels, respectively. This system is closer to the Mott transition, which happens at $U=4$. The main effect of $\Delta_1$, besides the already noted change in the gap magnitude, is the clearly visible change in the spectral intensity. For the largest value of $\Delta_1=0.5$, the superconducting gap remains around the Fermi points of the lower band. 

Finally, in Fig. (\ref{fig:rys1b}), we show the evolution of the superconducting state with $\Delta_1$ in a doped Mott insulator obtained for $U=6$, greater than the critical value $U_c=4$, and for the same values of other parameters as in the corresponding panels of Fig. (\ref{fig:rys1a}). Interestingly, the effect of CH interaction resulting in a change of $\Delta_1$ is hardly visible. The magnitude of the gap diminishes only very slightly from its value $\Delta_0=0.8$.
In the studied approximation and parameters' range there is no sign of superconducting correlation in the upper Hubbard band.
 
\section{Search of signatures of correlated hopping in N-QD-N structures}\label{sec:QD}

In quantum dots, electron energies are quantised and the strong interactions between them, described by the Hubbard term~\cite{glazman1988,ng1988}, are of primary importance. On the other hand, the presence of a correlated hopping term in nanostructures is expected on physical grounds as discussed in the Introduction. The Hubbard interaction $U$ leads to the appearance of the Kondo effect analogous to that observed in noble metals containing magnetic impurities; however, in nano-structures it manifests itself as a low-temperature increase of conductance, unlike in metals where it is visible as an increase of the resistance appearing with decreasing temperature. 

In systems with a quantum dot, the direct hopping between the dot and a metallic electrode is typically written~\cite{eckern2021} in a mixed representation as $V_{\lambda k \sigma}d_\sigma^\dagger c_{\lambda k \sigma}$, while the additional hopping-like interaction is written as $K_{\lambda,k,\sigma}d^\dagger_\sigma n_{\bar{\sigma}}c_{\lambda, k,\sigma}$, where operators $c_{\lambda, k,\sigma}$ describe annihilation of an electron with wave vector $k$, spin $\sigma$ in the electrode $\lambda$, while $d^\dagger_{\sigma}$ is the creation operator of a spin $\sigma$ electron on the quantum dot; $n_{\bar{\sigma}}=d^\dagger_{\bar{\sigma}} d_{\bar{\sigma}}$ is a number operator for spin $\bar{\sigma}=-\sigma$ electrons. The CH term has been earlier studied in ~\cite{meir2002,guinea2003,borda2004,stauber2004,lin2007,tooski2014,gorski2019,wrzesniewski2025,cronenwett2002,yu2005} mainly by numerical techniques. In a real situation, the value of correlated hopping $K$ may be of the order of the single-particle hopping $V$~\cite{dobry2011}. It may also be of the same or opposite sign and even depend on the relevant quantum numbers. Here, for the sake of simplicity, we assume that the ratio $x=-K_{\lambda,k,\sigma}/V_{\lambda,k,\sigma}$ is constant. The main questions we are interested in are: Is this term needed to describe, e.g., transport via a quantum dot? If so, how does it affect the results? How can one detect its presence experimentally?

The Hamiltonian of the N-QD-N system is written as 
\beq
{H}&=&\sum_{\lambda {k} \sigma}\varepsilon_{\lambda {k}}n_{\lambda {k} \sigma} + 
\sum _{\sigma} \varepsilon _{d\sigma}n_{\sigma} +Un_\uparrow n_\downarrow \nonumber \\
&+&\sum_{\lambda {k} \sigma} \left({V}_{\lambda {k}\sigma} c^{\dagger}_{\lambda k\sigma} D_{\sigma} 
+ {V}_{\lambda {k}\sigma}^* D^{\dagger}_{\sigma} c_{\lambda {k} \sigma}\right).
\label{eq:ham1}
\eeq
We use the same notation as previously~\cite{eckern2021,eckern2024}.  
The operators $c^{\dagger}_{\lambda k\sigma} (d^{\dagger}_{\sigma})$ create electrons in respective states $\lambda {k}\sigma$ $(\sigma)$ in the lead $\lambda$ (on the dot), while the operator $D_\sigma=d_\sigma(1-xn_{\bar{\sigma}})$ ($\bar\sigma = - \sigma$) takes care of the standard single - particle hopping with amplitude ${V}_{\lambda k\sigma}$ and the correlated hopping with amplitude $K_{\lambda k \sigma}$ {\it via} the parameter $x= - {K_{\lambda k \sigma}}/{V_{\lambda k \sigma}}$, assumed to be constant.

As the details of calculations have been presented in the cited papers, we shall not repeat them here. We only add that the effective couplings between the dot and the left/right electrode $\Gamma_{L/R}=2\pi\sum_{{k}}|V_{L/R{k}\sigma}|^2\delta(E-\varepsilon_{L/R{k}})$ are assumed to be constant (wide-band limit) and spin-independent. We use the Keldysh non-equilibrium GF approach to calculate the charge current $I(V,T,\Delta T)$ through the N-QD-N structure. The charge current depends on the voltage bias $V$, i.e.\ the difference between the chemical potentials of the left and right electrodes; $T$ is the overall temperature of the nano-structure, while $\Delta T$ is the temperature difference between both electrodes. In the calculations presented here, we shall assume a symmetric distribution of voltages ($V_{L/R}\pm V/2$) and no temperature differences $i.e.$ $T_{L/R}=T$. 

Typically, in transport experiments, one changes the position of the dot energy $\varepsilon_d$ by tuning the back-gate voltage. We are interested in the strongly non-linear transport with finite values of source-drain voltage $V$. From the knowledge of the current flowing through the system, we calculate the differential conductance $G_d (V)=\left({\partial I(V)}/{\partial V}\right)_{\Delta T=0}$ and the differential Seebeck coefficient~\cite{eckern2024} $S_d=\left(\frac{\partial V}{\partial \Delta T}\right)_{I} 
=-\left(\frac{\partial I}{\partial \Delta T}\right)_V/\left(\frac{\partial I}{\partial V}\right)_{\Delta T}$ by numerical differentiation.

\begin{figure}
\includegraphics[width=0.47\linewidth]{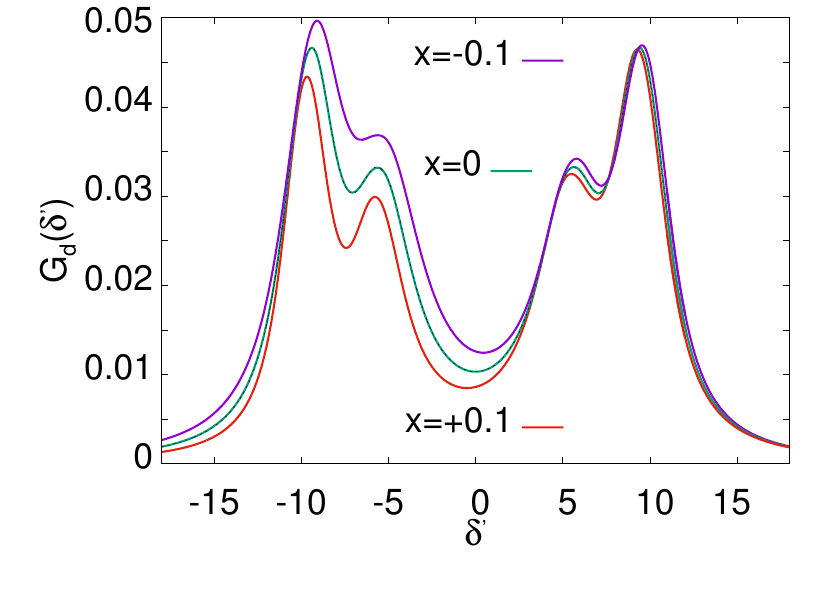}
\includegraphics[width=0.47\linewidth]{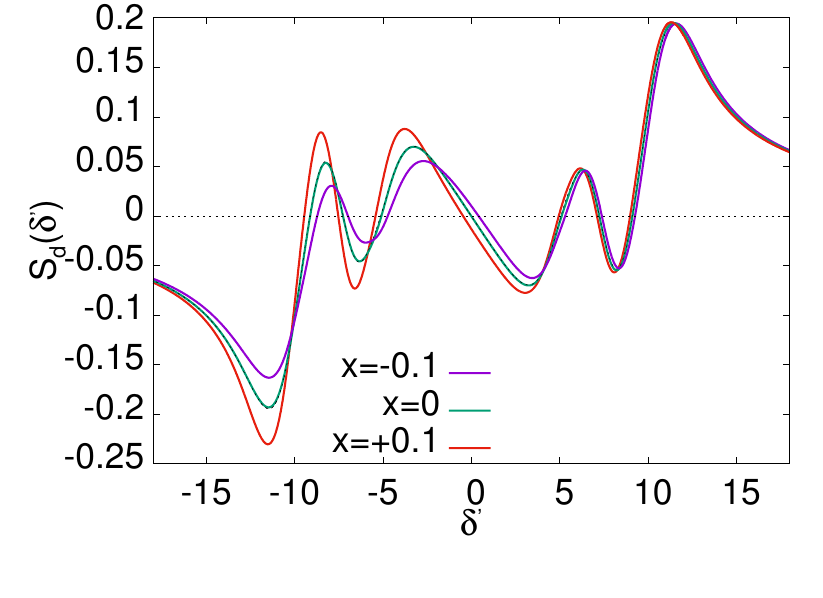}
\caption{
The differential conductance $G_d$ (left panel) and the differential Seebeck coefficient are plotted as functions of the back-gate voltage characterised by $\delta^\prime$ chosen in such a way as to show particle-hole symmetry for $x=0$. The other parameters are: Hubbard interaction $U=16$, temperature $T=0.3$, and bias voltage $V=4$ distributed symmetrically. All the parameters are expressed in units of the effective couplings to the electrodes $\Gamma_L=\Gamma_R=1$. Correlated hopping- characterised by $x$ - modifies mainly the lower Hubbard band, leaving the upper one nearly intact. } 
\label{fig:rys2}  
\end{figure}

In Fig. (\ref{fig:rys2}), on the left panel, we show the differential conductance as a function of the parameter $\delta'=\varepsilon_d+U/2+V/4$ chosen in such a way that the curve for $x=0$ is symmetric with respect to $\delta'=0$. The non-zero values of $x$ break the particle-hole symmetry and severely modify the currents. Importantly, only the part of the spectrum corresponding to the lower Hubbard sub-band is modified by the CH interaction. The upper Hubbard subband remains nearly unchanged with respect to $x=0$. Thus, the asymmetry of the conductance curve is a clear indication of the CH interaction. Moreover, for positive $x$ the conductance of the lower Hubbard band is smaller than for $x=0$, while for negative $x$ values, the differential conductance in the same region increases. As a result, the dip between the voltage-split maxima gets deeper for positive $x$, and it is shallower for negative $x$. 

These changes in the line shape of the differential conductance have their counterparts in the same dependence of the Seebeck coefficient, as visible in the right panel of Fig. (\ref{fig:rys2}). Again, the low-bias part ($\delta' < 0$) of the figure is severely modified, while the high-bias part is only very weakly affected. The thermopower calculated for $x=0$ is an anti-symmetric function of the (properly scaled) gate voltage. These changes of thermopower can be qualitatively understood from the Cutler-Mott relation, which states that thermopower can be obtained from conductance by differentiating the latter with respect to the chemical potential~\cite{cutler1969}.

\section{Summary and Conclusions}\label{sec:sum}

 In the present paper, we analyse the correlations beyond the standard Hubbard model in two different situations: the superconducting systems close to the Mott–Hubbard transition and nanostructure systems with quantum dots. In both cases, we have been interested in the role of correlated hopping, which is an interaction responsible for the occupation dependence of the hopping amplitude. In both physical situations studied here, CH breaks the particle–hole symmetry.

This symmetry breaking manifests itself as a different dependence of the superconducting temperature on whether the material is doped with electrons or holes~\cite{mmwysokinski2017}. The novel aspect of this work is the analysis of the CH interaction inducing a superconducting phase transition in materials. which undergo the Mott metal–insulator phase transition. This is achieved with help of the SHK model. We have used the Nambu–Gorkov Green function technique to solve for the superconducting state of the material close to the Mott transition, assuming that the CH interaction is a source of superconducting instability. The particle–hole symmetry-breaking term leads to changes in the superconducting gap visible in the spectral function. Other consequences will be studied in the near future. 

The CH interaction modifies the transport characteristics of N–QD–N structures. The symmetry-breaking character of CH in these circumstances manifests itself in the different modifications of lower and upper Hubbard sub-bands. We have found that the CH term in the Hamiltonian modifies the differential conductance only in the lower Hubbard sub-band. Without it, the conductance would be a symmetric function of the gate voltage. An experiment, carefully designed to study the symmetry of the non-linear conductance with respect to gate voltage, could identify the existence of this term and, moreover, its sign. The latter follows from the simple observation that the conductance in the lower Hubbard band increases for a negative sign of $x$ and decreases for positive $x$.

\section*{Acknowledgements}
It is our great pleasure and privilege to dedicate this work to Professor J\'ozef Spa\l{}ek on the occasion of his Jubilee of 50 years of scientific activities and his 80th birthday.

The work of KIW has been supported by the National Science Center, Poland (``Weave'' programme) through grant no. 2022/04/Y/ST3/00061. 
The research of MMW was supported by the “MagTop” project (FENG.02.01-IP.05-0028/23) carried out within the “International Research Agendas” programme of the Foundation for Polish Science co-financed by the European Union under the European Funds for Smart Economy 2021-2027 (FENG).

\end{document}